\documentclass[prb,twoside,twocolumn,showpacs,superscriptaddress,floatfix]{revtex4}
\usepackage{graphicx}
\usepackage{amsmath}
\usepackage{amssymb}

\newcommand{\YbPt}{Yb$_6$Pt}
\newcommand{\cpd}{Yb$_3$Pt$_4$}
\newcommand{\pupd}{Pu$_3$Pd$_4$}
\newcommand{\ab}{\emph{ab}-plane}
\newcommand{\ca}{\emph{c}-axis}
\newcommand{\sg}{\emph{R}$\overline{3}$}
\newcommand{\kvec}{{\bfseries{\emph{k}}}=0}
\newcommand{\mub}{$\mu_{\mathrm B}$}
\newcommand{\TN}{$T_\mathrm{N}$}

\begin{document}

\title{Determination of the magnetic structure of \cpd: \\ a \kvec\ local-moment antiferromagnet}
\author{Y. Janssen} \email[Email: ]{yjanssen@bnl.gov}
\author{M. S. Kim}
\affiliation{Condensed Matter Physics and Materials Science Department,
 Brookhaven National Laboratory, Upton, New York 11973}
\author{K. S. Park}
\affiliation{Condensed Matter Physics and Materials Science Department,
 Brookhaven National Laboratory, Upton, New York 11973}
\affiliation{Department of Physics and Astronomy, Stony Brook University, Stony Brook, New York 11794}
\author{L. S. Wu}
\author{C. Marques}
\author{M. C. Bennett}
\affiliation{Department of Physics and Astronomy, Stony Brook University, Stony Brook, New York 11794}
\author{Y. Chen}
\author{J. Li}
\affiliation{NIST Center for Neutron Research, NIST, Gaithersburg, Maryland 20899}
\affiliation{Department of Materials Science, University of Maryland, College Park, Maryland 20742}
\author{Q. Huang}
\author{J .W. Lynn}
\affiliation{NIST Center for Neutron Research, NIST, Gaithersburg, Maryland 20899}
\author{M. C. Aronson}
\affiliation{Condensed Matter Physics and Materials Science Department,
 Brookhaven National Laboratory, Upton, New York 11973}
\affiliation{Department of Physics and Astronomy, Stony Brook University, Stony Brook, New York 11794}
 
\date{\today} 
\begin{abstract}

We have used neutron diffraction measurements to study the zero-field magnetic structure
of the intermetallic compound \cpd,
which was earlier found to order antiferromagnetically at
the N\'{e}el temperature \TN=2.4~K, and displays a field-driven 
quantum critical point at 1.6 T. In \cpd, the Yb
moments sit on a single low-symmetry site in the rhombohedral lattice with
space group \sg.
The Yb ions form octahedra with edges that are twisted
with respect to the hexagonal unit cell, a twisting that results in every
Yb ion having
exactly one Yb nearest neighbor. Below
\TN, we found new diffracted intensity due to a \kvec\
magnetic structure.
This magnetic structure was compared to
all symmetry-allowed magnetic structures, and was subsequently refined.
The best-fitting magnetic structure model is antiferromagnetic, and
involves pairs of Yb nearest
neighbors on which the
moments point almost exactly towards each other.
This structure has moment components within the \ab\ as well as
parallel to the \ca, although the easy magnetization direction lies in the \ab. Our
magnetization results suggest that besides the crystal-electric-field
anisotropy, anisotropic 
exchange favoring alignment along the \ca\ is responsible for the overall direction of the ordered 
moments. The magnitude of the ordered Yb moments in \cpd\ is 0.81~\mub/Yb at 1.4~K.
The analysis of the bulk properties, the size of the ordered moment, and 
the observation of well-defined crystal-field levels argue that the Yb 
moments are spatially localized in zero field.
\end{abstract}

\pacs{75.25.+z, 71.27.+a, 75.20.Hr}

\maketitle

\section{Introduction}
\label{intro}

The quantum critical state 
in stoichiometric Yb intermetallic compounds is often associated with anomalous metallic states, 
non-Fermi liquid behavior~\cite{Lohneysen07, Stewart01, Stewart06, Gegenwart08}, 
and, most recently, with heavy-Fermion (HF) superconductivity~\cite{Nakatsuji08}.
Some stoichiometric Yb antiferromagnets show field-induced quantum critical points (QCP) 
and can thus be tuned towards quantum criticality by applying 
a magnetic field~\cite{Stewart01, Stewart06}. 
Different scenarios~\cite{Gegenwart08} can be related to
antiferromagnetic (AF) QCPs either involving large localized-moment
order or involving small-moment spin-density wave-like order. 
We have recently~\cite{Bennett08,Bennett09} 
identified the binary intermetallic compound \cpd\ as 
a stoichiometric Yb-based compound with a field-induced 
quantum critical point (QCP).

We have characterized \cpd\ by measuring  
thermodynamic and transport 
properties~\cite{Bennett08, Bennett09}.  
The magnetic susceptibility $\chi$ of \cpd\ 
above 200 K is well-described by a Curie-Weiss 
expression involving the full Yb$^{3+}$  moment of 4.5 $\mu_{B}$.
\cpd\ orders antiferromagnetically at the N\'{e}el temperature
\TN=2.4~K, releasing $\sim0.8R\mathrm{ln}2$ of entropy, consistent with largely 
localized Yb moments. 
Since the local symmetry of the Yb ions is triclinic,
the manifold of crystal field split states consists of four doublets. 
This scenario 
was confirmed in measurements of the specific heat $C$, which found that the 
ground doublet is well-separated from the first excited doublet. 
The resistivity below \TN\ is quadratic 
in temperature, the susceptibility $\chi$ weakly temperature dependent, 
and the Sommerfeld constant $\gamma = C/T$ vanishingly small. 
We concluded that the ordered state is a 
Fermi liquid, most likely weakly coupled to the ordered Yb moments. 
These experimental data indicate that the $f$-electrons are likely localized on the 
Yb ions, and are thus excluded from the Fermi surface. 
Besides this zero-field behavior, we have also shown that a field-induced quantum critical
point can be reached by applying fields of $\sim$1.6 T, and that 
the low-temperature high-field state is also a Fermi liquid.

Some physical parameters 
important to understand the properties of \cpd,
cannot be directly obtained from thermodynamic measurements, but need to
be determined by scattering methods. 
The magnetic propagation vector, which describes whether the magnetic structure
is commensurate or incommensurate with the crystal lattice,
is important for the underlying magnetic fluctuations and their coupling to quasiparticles.  
The order parameter, the temperature dependence of the 
ordered magnetic moment, can show whether the ordering 
transition is continuous or discontinuous.
The magnitude and direction of 
the ordered Yb moments, 
needed to rationalize the ordered-moment scenario (localized or itinerant), 
is determined by the magnetic structure.
We have used neutron diffraction experiments to determine 
all these, and the results are described in this paper.

This paper is organized as follows. We will give a 
description of the \cpd\ crystal structure that provides the groundwork for understanding 
the magnetically ordered structure. Then we will give results of
neutron diffraction experiments, and provide 
symmetry-allowed magnetic-structure models obtained by representation analysis. We 
will give our best-fitting magnetic-structure model. Finally, with the aid of
specific-heat, inelastic neutron scattering and single-crystal magnetization data,
we discuss the ordered Yb-4$f$ moments in terms of a local-moment picture.

\section{Experiments}

Small (up to $\sim$3 mg) single crystals of \cpd\ were grown out of a high-temperature ternary
solution\cite{Fisk89,Canfield92,Canfield01}, rich in Pb~\cite{Bennett09}.
A polycrystalline sample of \cpd\ was made by combining the yields of 12 different
growth crucibles. For initial characterization, we used a Philips diffractomer 
employing Cu-K$\alpha$ radiation to measure a powder x-ray
diffraction pattern
of finely ground crystals taken from each growth. The diffraction patterns were 
analyzed with the program
Rietica~\cite{rietica},
using a Le Bail-type~\cite{LeBail88} of refinement. As we will show below, they can be indexed
 according to the space group \sg,
with average lattice parameters
$a$=12.94(1)~\AA\ and $c$=5.655(5)~\AA.
These results are consistent with values $a$=12.888~\AA\ and $c$=5.629~\AA\
which were reported previously ~\cite{Palenzona77}.
No reflections from impurity phases were detected in any of the batches
used to make the neutron diffraction sample, which amounted to about
7~g. We note that we have measured the specific heat, ac susceptibility, and dc 
magnetization on crystals taken from many different growths and in no case have 
we found a detectable difference in Neel temperature or moment.
Additionally, we prepared a single crystal of about 60 mg and used it for a
single-crystal neutron diffraction experiment.

Neutron powder diffraction data were collected using the
high-resolution neutron powder diffractometer BT-1 and the
high-flux double focusing triple-axis spectrometer BT-7
at the NIST Center for Neutron Research NCNR. Samples were loaded
in a V (Al) can for the BT-1 (BT-7) experiment. In both experiments,
an ILL 'orange' cryostat was used, for full-pattern base-temperature
measurements
at $\sim$1.4~K, and for a full-pattern measurement at 5.1~K (5~K) in
the BT-1 (BT-7) experiment.
A Ge~311 (PG~002) monochromator produced neutrons with a wavelength
$\lambda$=2.079~\AA (2.359~\AA)
and data were collected over the angular range of
$2\theta$=3~-~168~$^\circ$ (5~-~60~$^\circ$)
with a step size of 0.05~$^\circ$ (0.1~$^\circ$) in the BT-1
(BT-7) experiment. Additionally, the 60 mg single crystal of \cpd was aligned 
such that a (110) reflection was in the Bragg condition,
and was subsequently used for a single-crystal neutron diffraction experiment 
conducted between 1.4~K
 and 4~K on the
triple-axis instrument BT-9 at NCNR-NIST.
The neutron diffraction data were analyzed by the Rietveld method using the
FULLPROF computer program suite~\cite{FULLPROF}. Representation analysis to
determine possible magnetic structures was performed using the computer
program SARAh~\cite{SARAh}.

Inelastic neutron scattering experiments were performed on the same 7~g
powder sample in the BT-7 double focusing triple-axis spectrometer with
 a fixed final energy of 14.7~meV, and at several fixed wave vectors and 
 several different temperatures both above and below T$_{N}$.
We used a double focusing PG monochromator, and a horizontally 
focusing analyzer to
obtain maximum intensity.

Specific heat between 0.5~K and 300~K, and magnetization at 4~K and in
fields up to 14~T were measured in Quantum Design Physical Property
Measurement Systems (PPMS) respectively equipped with a He-3 option
and with a vibrating sample magnetometer option.

\section{crystal and magnetic structure}

\subsection{Crystal structure}

\label{structure}


\begin{figure}[!ht]
\includegraphics[width=0.48\textwidth,trim=0in 0in 0in 0in]{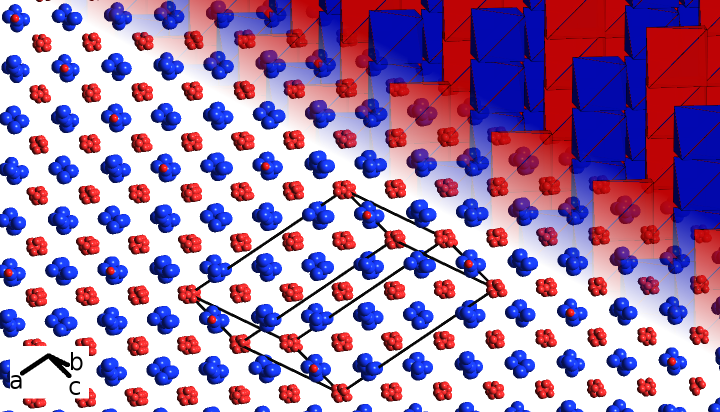}
\includegraphics[width=0.48\textwidth]{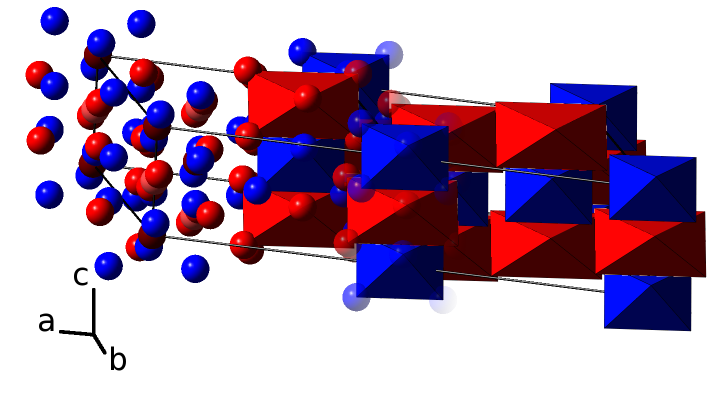}
\begin{center}
\caption{(Color online) Schematic drawing of the crystal structure of \cpd, produced 
with the software of Ozawa and Kang~\cite{Ozawa04}.
The top figure shows an unusual projection, normal to a (10~-30~9) plane,
showing how the rhombohedral crystal structure can be seen in relation
to a simple cubic structure.
Structural octahedra (bottom), as explained in the text, depict
'Pt$_6$Pt' (red) and 'Yb$_6$Pt' (blue). The relation between these
structural octahedra and the simple cubic structure is shown as well.}\label{Crystalfig}
\end{center}
\end{figure}

\cpd\ crystallizes in the rhombohedral \pupd-type of structure, where
Pu can be replaced by all rare earths except Eu, as well as by Y, and
Pd by Pt~\cite{Cromer73, Palenzona74, Iandelli75,Palenzona77}.
Th$_3$Pd$_4$~\cite{Palenzona74} and Zr$_3$Pd$_4$~\cite{Bendersky93}
have also been reported to form in this structure type. For the
rare-earth compounds, no deviations from the lanthanide contraction
have been observed~\cite{Palenzona74, Palenzona77}, indicating
that both Ce and Yb are trivalent in these compounds.

Although YbPt forms in the FeB-type structure~\cite{Johnson71},
not in the layered cubic CsCl-type structure, we can describe the
rhombohedral \cpd\ structure
in terms of this CsCl-type YbPt~\cite{Cenzual88}. Fig.~\ref{Crystalfig}
shows an unusual projection of
the crystal structure of \cpd, viewed normal to a (10~-30~9) plane.
There are layers of Pt atoms, and layers of
Yb atoms. In these Yb layers, 1/7 of the atoms is replaced by Pt atoms in an
ordered fashion.
Besides this, the \cpd\ structure is also distorted away from cubic symmetry, a
distortion which also involves a shortening of the distances between the
Yb-layer Pt atoms and their surrounding Yb atoms~\cite{Bendersky93}.
This shortening allows us, for descriptive purposes, to divide the
structure into two types of spatially separated octahedra with a Pt
atom in the center. 
Of these, one type has Yb atoms at $18f$ 
on the corners surrounding the Pt atom
at $2a$, and may be considered 'Yb$_6$Pt' octahedra.  
The other type has
Pt atoms at $18f$ surrounding the Pt atom at $2b$, and are thus
 'Pt$_6$Pt' octahedra. Note that these octahedra are not
coordination polyhedra, which can be found in
Refs.~\onlinecite{Bennett09, Cromer73}. 
In the top right part of
Fig.~\ref{Crystalfig} the structure is represented with these two types of octahedra.
Viewed in this unusual way, the \cpd\ crystal structure may be visualized
as a stacking of layers of 'Yb$_6$Pt' and 'Pt$_6$Pt' octahedra.
Note that the sides of the two different kinds of stacked octahedra are almost
parallel.


\begin{figure}[!ht]
\includegraphics[width=0.48\textwidth,trim=0in 0in 0in 0in]{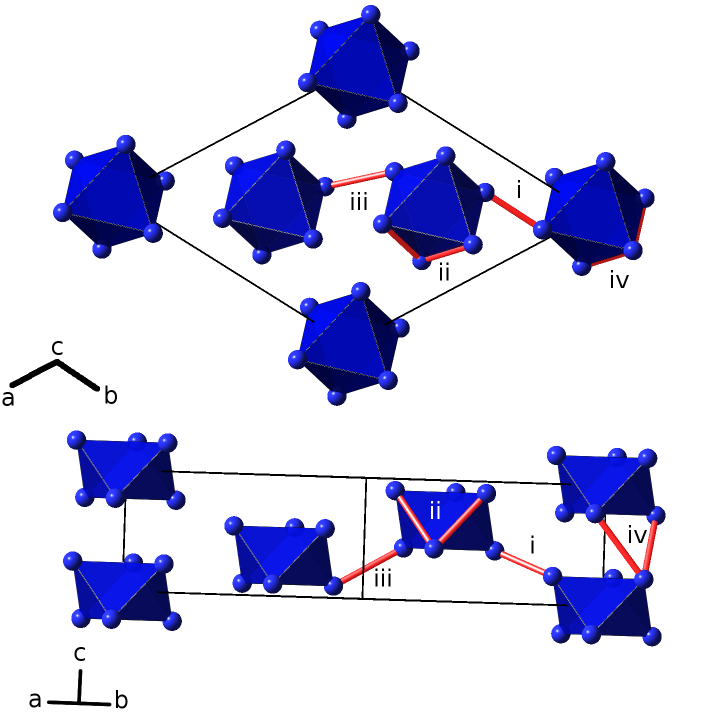}
\begin{center}
\caption{(Color online) Schematic drawings of structural Yb octahedra 
in the unit cell of
\cpd,  from two perspectives,
 emphasizing the Yb-Yb near-neighbor distances shorter
than 4~\AA, denoted as i, ii, iii and iv. Every Yb atom has one (i) Yb
nearest neighbor residing on a neigboring octahedron,
at $\sim$ 3.37 \AA, and two next-nearest Yb neighbors (ii) on the same
structural octahedron, at $\sim$ 3.64 \AA. Two pairs
of farther-neighboring Yb atoms lie at (iii) $\sim$3.80~\AA  and at
(iv) $\sim$3.89~\AA. }\label{YbYb}
\end{center}
\end{figure}

The rhombohedral crystal structure of \cpd\ can be represented by 
parallel chains
of alternating 'Yb$_6$Pt' and 'Pt$_6$Pt' octahedra.
The octahedra on neighboring chains are shifted with respect to one
another. A rhombohedral unit cell consists of a stacked
pair of these unequally sized, parallel-sided octahedra. A hexagonal
unit cell, see Fig.~\ref{Crystalfig}b, consists of
a stacked pair translated by (2/3,1/3,1/3) and by (1/3,2/3,2/3).
Though parallel to one another, the sides of the 'Yb$_6$Pt' and 'Pt$_6$Pt'
octahedra are not parallel to the crystal axes. Fig.~\ref{YbYb} shows 
a cross section of this hexagonal unit cell, showing the
Yb-Yb distances which are shorter than 4~\AA. Significantly, every Yb 
atom has only one Yb nearest neighbor, which greatly simplifies the 
magnetic structure, as we will see below.

\subsection{High-resolution neutron powder diffraction}
\label{hrpd}

\begin{figure}[!ht]
\includegraphics[width=0.48\textwidth,trim=0in 0in 0in 0in]{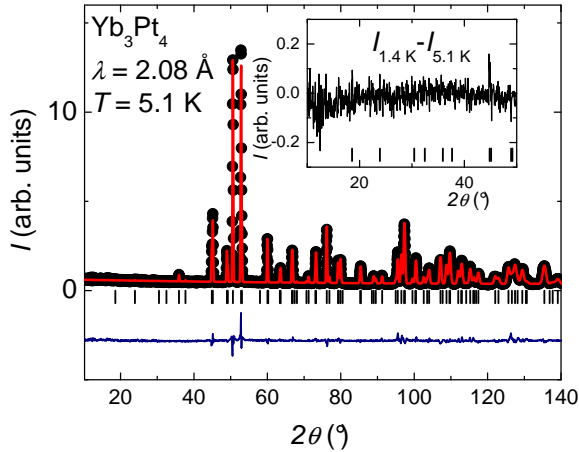}
\begin{center}
\caption{(Color online) High-resolution neutron powder diffraction data (dots),
Rietveld fit (line) and allowed Bragg reflections (tick marks) for
\cpd, measured at 5.1~K. The lower trace is the difference between the data and the fit, on the
same scale. The inset shows the difference between the 5.1~K data
and 1.4 K~data (line) indicating that the difference between these two
is statistically insignificant even at the indicated Bragg reflections (tick marks).
Error bars where indicated are statistical in origin and represent one standard deviation.}\label{hrnpd} 
\end{center}
\end{figure}

We present in Fig.~\ref{hrnpd} a high-resolution neutron powder 
diffraction pattern which was measured at 5.1~K, i.e. in the paramagnetic phase.
Rietveld analysis  indicates that the diffraction pattern can be
indexed according to the room-temperature space group \sg, with
lattice parameters $a=12.8687(2)$~\AA\ and $c=5.6160(1)$~\AA.
These lattice parameters are smaller than the room-temperature
values $a=12.94$~\AA\ and $c=5.655$~\AA\ reported above, which
can be ascribed to thermal contraction.

Further Rietveld refinement of the 5.1~K diffraction pattern fully conforms
to the \pupd -type of structure as described in Sec.~\ref{structure}.
The results of our best fit, which included background parameters,
a scale factor, and peak shape parameters, are shown in
Fig.~\ref{hrnpd} and are summarized in Table~\ref{npdstruct}.
In accordance with detailed work on single crystals~\cite{Bennett09},
we found that  variations of the site occupancies were insignificant,
and therefore we fixed the sites to be fully occupied.
The refined coordinates for the Yb-$18f$ and the Pt-$18f$ are very
similar to the room temperature values found by Palenzona and, in more
 detail, by Bennett et al.~\cite{Palenzona77, Bennett09}.
Including isotropic thermal factors improved the fit. Our best fit
was made with different thermal factors for the different species,
Yb and Pt. Permitting the possibility of different thermal factors for the different
 Pt sites provided no further improvement to the fit.
We also included a preferential-orientation parameter to improve
our refinement, using a modified March function, as implemented
by Fullprof~\cite{FULLPROF}. The modeled preferred-orientation
vector was parallel to the \ca, and the refined parameter was
found equal to 0.921(2), indicating a slight needle-like habit of the crystals. This
is reasonable since the crystals that we coarsely ground for the powder
typically were somewhat longer in the $c$-direction than in the
planar directions.


\label{diffraction}
\begin{table}[h]
\begin{center}
\begin{tabular}{cccccc}
\hline\hline Atom(site) & \textit {x/a} & \textit {y/b} & \textit
{z/c}  & occ.& B$_\mathrm{iso}$ \AA  \\
\hline
Pt($3a$) & 0 & 0 & 0 & 1 & 0.28(5) \\
Pt($3b$) & 0 & 0 & 0.5 & 1 & 0.28(5) \\
Pt($18f$) & 0.2693(2)& 0.2171(2) & 0.2786(4) & 1 & 0.28(5) \\
Yb($18f$) & 0.0421(2)& 0.2111(1) & 0.2335(3) & 1 & 0.47(4) \\
\hline\hline
\end{tabular}
\caption{Refined structural parameters for \cpd\ obtained from
high-resolution neutron diffraction at 5.1~K. Space group \sg, 
\emph{a}=12.8687(2)~\AA, \emph{c}=5.6160(1)~\AA, 
\emph{R}$_\mathrm{p}$=~0.055, \emph{R}$_\mathrm{wp}$=~0.072, 
$\chi^2=3.8$.
\label{npdstruct}}
\end{center}
\end{table}

Having used the data taken at 5.1~K to verify the crystal structure of our powder sample, 
we next measured a neutron powder diffraction pattern at 1.4~K, i.e in the magnetically ordered phase below
\TN=2.4 K. The inset of Fig.~\ref{hrnpd}
shows the difference between the two diffraction patterns. This comparison did 
not reveal any magnetic intensity, finding that there is no statistically significant 
difference between the diffraction intensities obtained above and below T$_{N}$.
A separate refinement of the low temperature diffraction pattern established that 
there is no shift of the Bragg peak positions, and that the lattice parameters refined from the
5.1 K and the 1.4 K data were the same within $1\cdot 10^{-4}$. As we will show in 
the next section, magnetic intensity was only observed when we sacrificed resolution 
in favor of scattered neutron intensity.

\subsection{High-intensity neutron powder diffraction}


\begin{figure}[!ht]
\includegraphics[width=0.48\textwidth,trim=0in 0.5in 0.5in 0in]{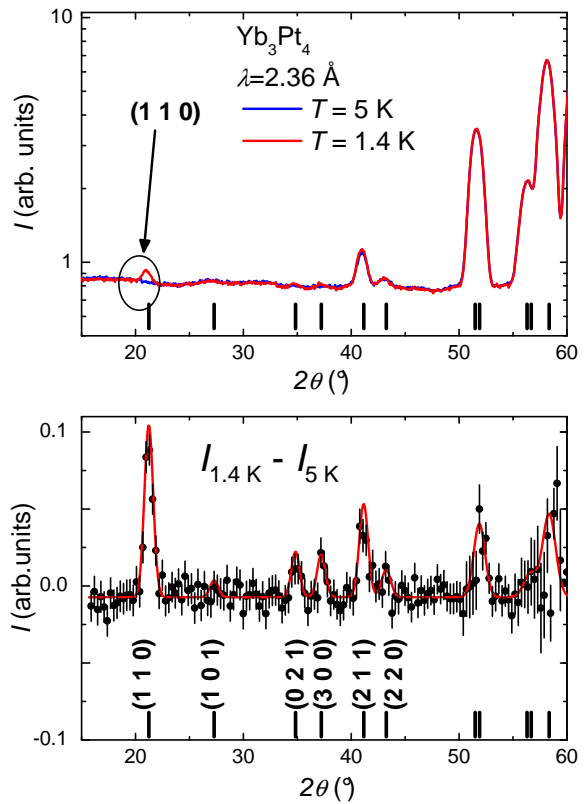}
\begin{center}
\caption{(Color online) (Top) Low-resolution neutron powder diffraction
patterns (log scale), measured at 5~K and at 1.4~K. The tick marks
indicate Bragg positions. The lines connecting the data points
are guides for the eye. The arrow indicates the strongest
magnetic Bragg-peak position, (110). (Bottom) The difference between the 
neutron
powder diffraction patterns at 5~K and 1.4~K.  (linear scale). The (red) line is a
profile-matched Rietveld fit, described in the text. Bragg-peak positions are indicated by
tick marks. The indexed peaks were used for magnetic structure
refinement.}\label{lrnpd}
\end{center}
\end{figure}

We successfully observed new magnetic intensity below \TN\
in high-intensity-low-resolution powder diffraction experiments.
Fig.~\ref{lrnpd}a compares diffraction patterns
measured at 5 K and at 1.4 K, demonstrating that the (110) reflection
is significantly more intense at 1.4 K than at 5 K. Fig.~\ref{lrnpd}b
shows the difference between patterns measured at 5 K and at 1.4 K.
This difference pattern shows that at 1.4 K, there is new diffracted
intensity at multiple
Bragg-peak positions, making feasible the determination of the magnetic 
structure which is the major result we report here.


\begin{figure}[!ht]
\includegraphics[width=0.48\textwidth,trim=0in 0in 0in 0in]{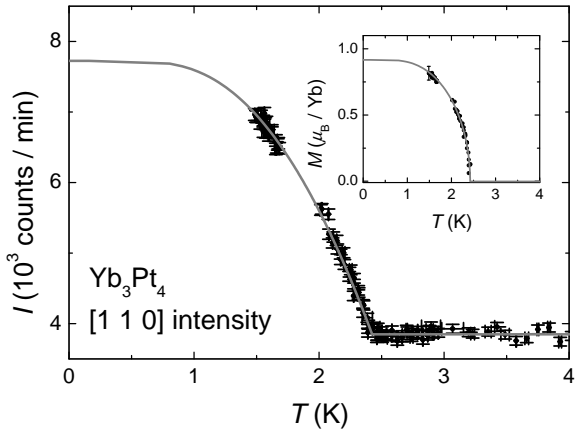}
\begin{center}
\caption{The temperature dependent intensity of the (110) nuclear and magnetic Bragg peak measured on a
single crystal. The was line was calculated from the mean-field spin $\frac{1}{2}$
model with \TN= 2.42(3) K. The inset shows the
 temperature-dependent ordered Yb moment, where the overall scale factor was determined from
 refined 1.4 K diffraction data as described in the text.}\label{OPmeasured}
\end{center}
\end{figure}

We confirm that this new intensity is related to the antiferromagnetic (AF) 
transition by a performing single-crystal neutron diffraction experiment.
Fig.~\ref{OPmeasured} shows the temperature-dependent intensity
of the [110] reflection, measured by summing the counts of the
centered BT9 detector, at temperatures between 1.4 K and 4 K.
The data were obtained over the course of multiple 
sweeps, with temperatures both increasing and decreasing. Starting
at the lowest temperature, this [110] intensity decreases
continuously, up to about 2.4 K, above which it remains
constant. These results indicate a magnetic ordering
temperature of about 2.4 K, in excellent agreement with our
previous results from specific heat, dc and ac magnetic susceptibility, 
and electrical resistivity.~\cite{Bennett09,Bennett08}.
As was also the case in these earlier measurements,  we observed neither thermal
hysteresis nor a sudden change in this order parameter near T$_{N}$, suggesting that the
ordering transition is continuous and well-described by a mean-field expression.
Furthermore, because the magnetic intensity in every case appears at Bragg-peak
positions which have intensity above \TN,
the magnetic unit cell is the same as the crystallographic unit
cell. We note that this is also the case~\cite{Bonville94} for the magnetic structure 
of isostructural Yb$_3$Pd$_4$. Such a structure, with propagation vector \kvec, is
often associated with ferromagnetism, but in this case, as
we will see below, it is due to a \kvec\ antiferromagnetic structure.

\subsection{Representation analysis}

We will test our diffraction data against magnetic-structure models
obtained from representation analysis.
These models are linear combinations of magnetic structure
patterns, called basis vectors (BVs), for magnetic-moment carrying atoms.
The choice of BVs to be included in a magnetic structure model is
limited, and is wholly determined by symmetry. A number of BVs form an 
irreducible representation (IR)
or a corepresentation (CR) of those space group elements that
leave the magnetic propagation vector {{\bfseries{\emph{k}}}} invariant.
If we apply Landau theory for continuous phase transitions then
only one representation, an IR or a CR~\cite{Wills08, Radaelli07},
is involved in the magnetic order, and therefore the BVs belonging
to a single representation.
Since our  magnetic propagation vector $k=0$, we need to examine
 the BVs belonging to the IRs and CRs of the space group \sg\ itself,
for the magnetic Yb atoms on the \YbPt\ octahedra.
According to the program SARAh, there are six one-dimensional IRs, with
three BVs each. Among these six IRs, there are two pairs of
CRs~\cite{Tinkham92, Wills08}, namely IRs $\Gamma3$ and $\Gamma5$,
and $\Gamma4$ and $\Gamma6$, which means that these IRs, as well
as their BVs, will be considered combined.


\begin{figure}[t]
\includegraphics[width=0.48\textwidth,trim=0in 0in 0in 0in]{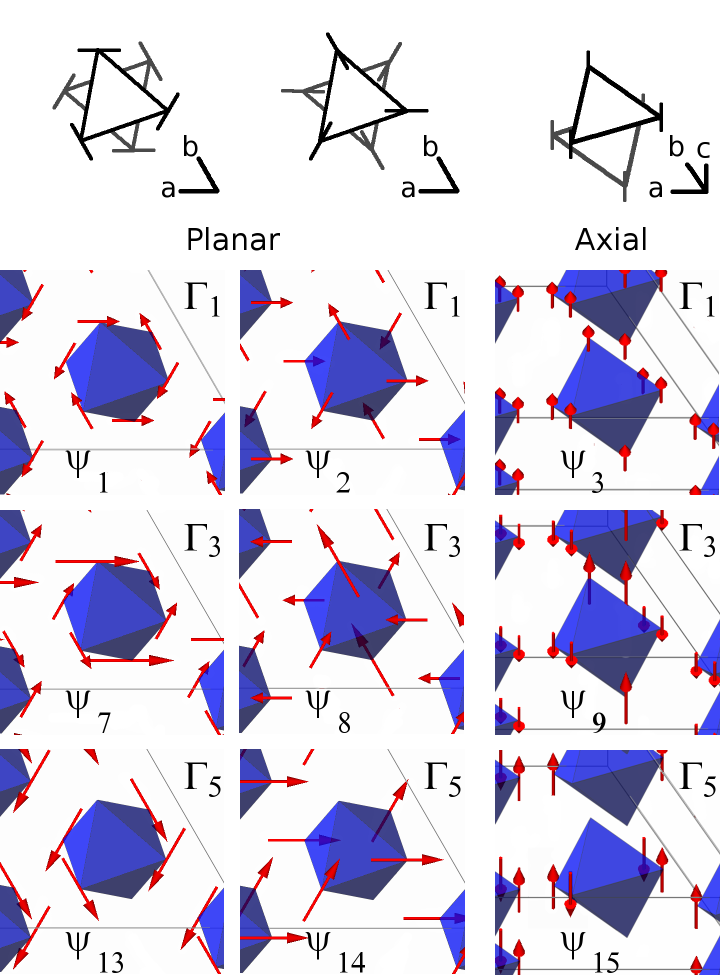}
\begin{center}
\caption{(Color online) Basis vectors belonging to 'ferromagnetic' IRs $\Gamma$1,
$\Gamma$3 and $\Gamma5$, used to model the magnetic structures for \cpd.
}\label{FIRs}
\end{center}
\end{figure}


\begin{figure}[t]
\includegraphics[width=0.48\textwidth,trim=0in 0in 0in 0in]{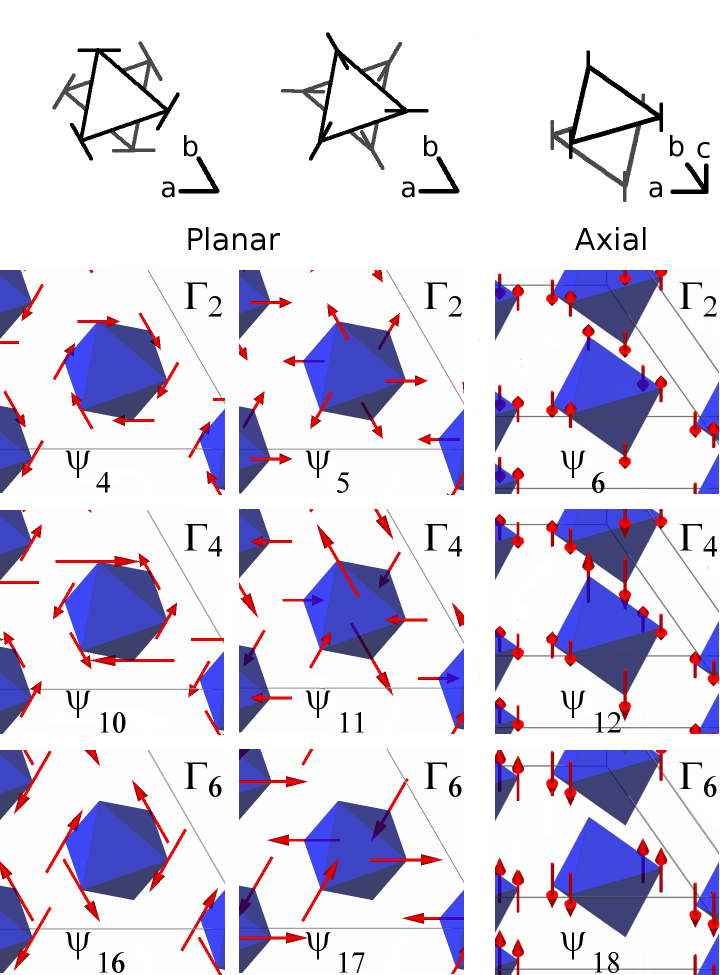}
\begin{center}
\caption{(Color online) Basis vectors belonging to 'antiferromagnetic' IRs $\Gamma$2,
 $\Gamma$4 and $\Gamma6$, used to model the magnetic structures for \cpd.
}\label{AFIRs}
\end{center}
\end{figure}

Each IR has two BVs that describe components of the magnetic structure
in the plane perpendicular
to the \ca\ (from here: planar), and one BV that describes a component
 of the magnetic structure parallel to the \ca\ (from here: axial).
The IRs can be split in two parts, $\Gamma n$ ($n=1,3,5$), and
 $\Gamma n$ ($n=2,4,6$), respectively.
For $n=1,3,5$ \emph{some} BVs show a non-zero net magnetization,
and are ferromagnetic.
For $n=2,4,6$ \emph{all} BVs show no net magnetization, and are
antiferromagnetic.
A schematic drawing of the 18 BVs is given in Fig.~\ref{FIRs}, for
the IRs with (a) ferromagnetic BV(s), and Fig.~\ref{AFIRs}, for the
antiferromagnetic IRs, respectively. Every hexagonal unit cell
contains three \YbPt\ structural octahedra, and on each of these
the magnetic-moment configuration is the same. Therefore, we
choose to display the BVs centered on a single \YbPt\ octahedron,
emphasizing the
planar moments by viewing octahedra parallel to the \ca, and
emphasizing the axial moments by viewing octahedra at an angle of
$\sim$ 30$^\circ$
from the \ca.
In the figures, each row contains a different IR. In each, the third
column contains the axial BVs, while the first two columns contain the planar BVs.
The IRs $\Gamma_n$ can also be divided in three categories:
two IRs ($n=1,2$) with BVs that have equal moments on all Yb sites,
two IRs ($n=3,4$) with BVs that have 2/3 of the Yb moments half the
size of the other 1/3,
and two IRs ($n=5,6$) with BVs that have non-zero size on 2/3 of the
Yb moments and zero on the other 1/3.

A structural octahedron consists of two equilateral triangles with moments on each vertex.
For the equal-moment configurations with the moments planar, these
triangles carry AF
structures where the moments on the vertices are aligned with the
crystal axes and rotated by 120$^\circ$ with respect to one another.
For the non-equal-moment configurations with the moments planar,
these triangles form  canted F structures.
For the equal-moment configurations with axial moments, these
triangles correspond to F structures, but if the moments on each site are \emph{not} equal then
the structures on these triangles are AF.
Pairs of IRs $\Gamma$1 and $\Gamma$2, $\Gamma$3 and $\Gamma$4,
$\Gamma$5 and $\Gamma$6, are analogous, as their respective BVs form
pairs which are composed of the
same triangular moeities described here. The difference between the BVs of
the 'F' and the 'AF' IRs, is the direction of the moments on the
top triangle relative to those on the bottom triangle, as can be appreciated 
by comparing, for example, BV $\psi 1$ ($\psi 2$)  to $\psi 4$ ($\psi 5$).

Figs.~\ref{FIRs} and \ref{AFIRs} show that the moments on individual Yb 
sites are rotated by 120$^\circ$ in the two planar BVs comprising an IR.
However, it is not possible to rotate the crystal and at the
same time obtain the other planar BV belonging to that IR.
We are therefore able to distinguish between different planar
BVs for a given IR from a powder diffraction
experiment~\cite{Shirane59,Wills08}.
We note further that the IRs which are part of a CR here each
produce the same structure factors in our calculations, and
therefore $\Gamma$3(4) cannot be distinguished from $\Gamma$5(6). 
We will therefore only use IRs $\Gamma3$ and $\Gamma4$, as well as 
$\Gamma1$ and $\Gamma2$ for our refinements.

In a recent publication~\cite{Litvin08}, Litvin
produced crystallographic tables of magnetic space groups. The magnetic
space groups presented here are
based on the so-called two-colored space groups, and consequently only include equal-moment
magnetic structures. Equal-moment IRs $\Gamma1$ and $\Gamma2$ are
comparable to the magnetic space groups \sg\ and
\emph{R}$\overline{3}$', respectively.
We note that models which can be constructed using the unequal-moment
IRs $\Gamma3-6$ cannot be constructed from these magnetic
space groups, requiring instead a more involved multicolored space-group
analysis~\cite{Harker81}.

\subsection{Magnetic structure refinement}

The 1.4 K low-resolution diffraction pattern presented in Fig.~\ref{lrnpd}\textbf{a}
was used for a Le Bail-type~\cite{LeBail88} profile match, fitting for
zero-offset, lattice parameters, background, and peak-shape parameters.
For the refined lattice parameters we found $a=12.803(7)$ \AA, and
$c=5.602(4)$ \AA, in agreement with values found in
our high-resolution experiment described above. The agreement factors
for this fit were $R_\mathrm{p} = 0.019$ and $R_\mathrm{wp} = 0.024$.

We determined a scale factor from the 5 K diffraction pattern of
Fig.~\ref{lrnpd}. For this, we use the profile-matched parameters
 found for the 1.4 K pattern,
together with the atomic positions, thermal factors, as well as
the preferred-orientation parameter, found from the refinement
of the 5.1 K high-resolution data of Sec.~\ref{hrpd}. This
yielded the scale factor with a refined estimated error of 0.4\%.
 The agreement factors for this fit were $R_\mathrm{p} = 0.027$
and $R_\mathrm{wp} = 0.037$, which compares well to the values
obtained from the profile matching above.

\begin{figure}[!ht]
\includegraphics[width=0.48\textwidth,trim=0in 2.5in 0in 0in]{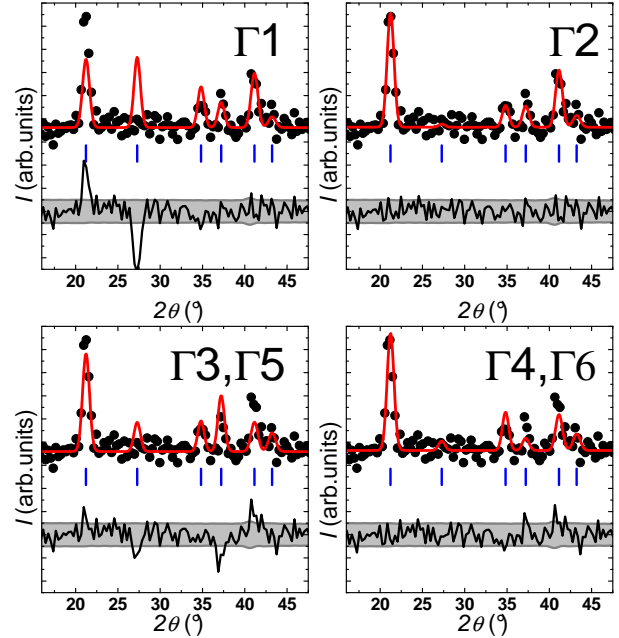}
\begin{center}
\caption{(Color online) For each of the considered IRs
($\Gamma1-4$): Magnetic-only diffraction pattern (dots)
for \cpd\ at 1.4 K, together with best fit (line) and difference
pattern (line) on top of shaded error band estimated from
data of Fig.~\ref{lrnpd}.}\label{fitcompare}
\end{center}
\end{figure}

We determined the best fitting magnetic-structure model for each
of the four IRs under consideration by refining the BV coefficients for the
'magnetic-only' subtracted pattern of Fig~\ref{lrnpd}b while we kept
the scale parameter and the other parameters determined above (except
the background parameters), fixed to the value determined on the
full 1.4 K and 5 K patterns. We included diffracted intensity
only up to 2$\theta$ = 45$^\circ$, thus avoiding the relatively
noisy background caused by the strong nuclear peaks at
larger diffraction angles in Fig.~\ref{lrnpd}\textbf{b}.
An initial Le Bail-type profile refinement yielded agreement
factors of  $R_\mathrm{p} =0.0142$ and $R_\mathrm{wp} =0.0177$.
This can be considered the best possible fit,
since here the Bragg-intensities themselves are fit parameters,
and generated to best match the observed intensities~\cite{LeBail88}.
The results of the refinements of
the best-fitting magnetic structure models are shown in
Fig.~\ref{fitcompare}, and summarized in Table~\ref{magstruct}.
The overall best-fitting magnetic structure model is
generated by the BVs of the AF IR $\Gamma$2, since that model
gave \emph{exactly} the same agreement factors as the model-free
profile refinement. Second best is the AF
best-fitting model described by the BVs of IR $\Gamma$4. The
 the best-fitting model generated by 'F'
IRs $\Gamma$3 and $\Gamma$1 provides a significantly poorer description 
of the magnetic diffraction, and it should be
noted here that the refinement for $\Gamma$1 was not stable.

Table~\ref{magstruct} also includes the size of the ordered Yb magnetic
moment for the best-fitting models.
Averaged moments are given for the structures belonging to unequal-moment IRs
$\Gamma$3 and $\Gamma$4. The average
ordered moments for the four best-structure models due to $\Gamma$1,
 $\Gamma$2, and $\Gamma$4 are all
close to 0.85 $\mu_\mathrm{B}$, whereas the best-fitting model
belonging to IR $\Gamma$3 has an average moment of 1.6 $\mu_\mathrm{B}$/Yb.
The ordered moment for the best-fitting structure is about 0.81(5)
$\mu_\mathrm{B}$/Yb at 1.4 K. This value was used, together with
the temperature-dependent intensity of
the single-crystal [110]-reflection, to normalize the order parameter,
which is displayed in the inset of Fig.~\ref{OPmeasured}.

Finally, we refined the scale factor for the full 1.4 K diffraction
pattern, including also the best magnetic structure model according
to IR $\Gamma$2.
Within error, we found the same scale factor as the one we used for
 the magnetic structure refinement,
with agreement factors of $R_\mathrm{p} =0.025$ and $R_\mathrm{wp} =0.036$,
comparable to the agreement factors found for the 5 K data,
as described above.


\begin{table}[h]
\begin{center}
\begin{tabular}{cccccc}
\hline\hline IR & BV & coefficient& $\mu$ ($\mu_\mathrm{B}$/Yb) &
R$_\mathrm{p}$ & R$_\mathrm{wp}$ \\
\hline
$\Gamma$1 & $\psi$1 & 0.44(4) &            &        &        \\
          & $\psi$2 & 0.36(2) &            &        &        \\
          & $\psi$3 & 0.34(3) &    0.89(9) & 0.0233 & 0.0403 \\
\hline
$\Gamma$2 & $\psi$4 & 0.38(1) &            &        &        \\
          & $\psi$5 & 0.34(2) &            &        &        \\
          & $\psi$6 & 0.18(2) &    0.81(5) & 0.0142  & 0.0177\\
\hline
$\Gamma$3 & $\psi$7 & -0.11(4) &            &        &        \\
          & $\psi$8 & -0.23(2) &            &        &        \\
          & $\psi$9 &  0.247(8)&    1.6(2) & 0.0168  & 0.0204 \\
\hline
$\Gamma$4 & $\psi$10& 0.14(2) &            &        &        \\
          & $\psi$11& 0.08(7) &            &        &        \\
          & $\psi$12& 0.228(5)&    0.86(9) & 0.0152 & 0.0189 \\
\hline\hline
\end{tabular}
\caption{Magnetic refinement results for \cpd\ at 1.4 K are tabulated for each of the considered
IRs, with their BVs and their best-fit coefficients, as well as the average
Yb moment $\mu$
and the agreement factors $R_\mathrm{p}$ and $R_\mathrm{wp}$  for each model.
The best-fitting model has $\Gamma$2.
\label{magstruct}}
\end{center}
\end{table}


\begin{figure}[!ht]
\includegraphics[width=0.48\textwidth,trim=0in 0in 0in 0in]{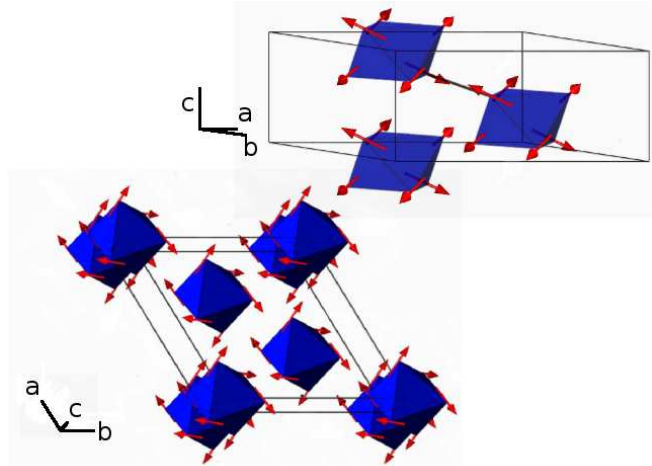}
\begin{center}
\caption{(Color online) Two perspectives of a schematic view of the best refined magnetic structure model
for \cpd\ at 1.4 K.}\label{Magstructure}
\end{center}
\end{figure}

A schematic drawing of the proposed magnetic structure for \cpd\ is
presented in Fig.~\ref{Magstructure}. In this structure, the ordered Yb
moments are all of the same size
and have both planar and axial components. Yb octahedra form the basic 
building block of the extended magnetic structure. In our model all Yb moments point
outward from such an octahedron, where the Yb moments on the top triangle
point upward, and the moments on the bottom triangle point downward.
The directions of the moments are such that the moments on two
nearest-neighbor Yb ions, residing in neighboring Yb octahedra, 
point almost exactly towards each
other, see Fig.~\ref{Magstructure}.
As calculated from Table~\ref{magstruct}, the planar component
of the ordered moment is 0.73~\mub, while the axial component
is 0.34~\mub, thus the ordered moments make
an angle of of 64$^\circ$ with the \ca.

\section{Crystal-field and exchange anisotropy}

\begin{figure}[!ht]
\includegraphics[width=0.48\textwidth,trim=0in 0in 0in 0in]{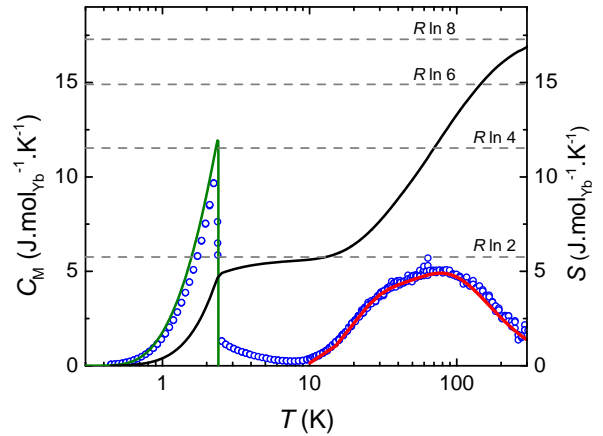}
\begin{center}
\caption{(Color online) The difference (open circles) between the measured specific heats 
of \cpd and Lu$_{3}$Pt$_{4}$, $C_\mathrm{M}$, and its associated entropy $S$ (solid black line). 
The green line is an idealized mean-field spin $\frac{1}{2}$ specific-heat peak with \TN=2.4~K.
The red line is a fit to a Schottky expression, yielding excited doublets at 7.1,
20, and 30 meV.}\label{CP}
\end{center}
\end{figure}
A central goal of our magnetic structure determination was to gain insight 
into whether the Yb moments should be considered spatially localized in 
zero field, or not. In this section, we seek to rationalize the magnitude 
of the moment deduced from our diffraction experiments within a crystal 
field model, including both the crystal field and exchange anisotropies.
In the triclinic environment of the Yb ions in \cpd, the
crystal-electric field (CEF) lifts the 2$J$+1 (= 8)-fold
degeneracy of the magnetic $4f$ shell into 4 doublets, causing
the single-ion CEF anisotropy.
As stated previously, specific heat measurements provided evidence 
for this crystal field splitting. We present in Fig.~\ref{CP} an improved 
analysis of the magnetic contribution to specific heat $C_\mathrm{M}$, 
obtained  by subtracting the 
Lu$_3$Pt$_4$~\cite{Lu3Pt4} specific heat from the measured \cpd\ specific 
heat $C(T)$. 
The ordering peak at \TN= 2.4 K~\cite{Bennett09} is prominent and looks 
similar to an included ideal mean-field $S=1/2$-specific heat 
peak~\cite{Ott80,Ott82}, but it reaches only about 80\% of the 
theoretical value at \TN. 
The entropy associated with magnetic order is very similar as in the 
earlier measurement, 
approaching the doublet value of $R$ln2 only at $\sim$ 10 K. We note that, 
in agreement with our earlier report, the Sommerfeld coefficient 
$\gamma=C/T$ in the ordered state is very small, certainly no larger than 
10 mJ/mol-K$^{2}$ at $T$=0 K. 
Also clear is a broad Schottky-like peak with a maximum near 75 K.
This peak was fit above 10 K with a Schottky-like
expression accounting for the ground-state doublet and three
excited doublets. The line in Fig.~\ref{CP} shows the quality of the fit with
these excited doublets at energies of 7.1 (5) meV, 20 (2) meV,
and 30 (6) meV. These values are notably different from values found by
Bennett et al.~\cite{Bennett09}, who found 4.3 meV and 10.9 meV, which
we mainly ascribe to fact that they used a Debye expression to account
for the lattice contribution of specific heat, where we used Lu$_3$Pt$_4$ data.

\begin{figure}[!ht]
\includegraphics[width=0.48\textwidth,trim=0in 0in 0in 0in]{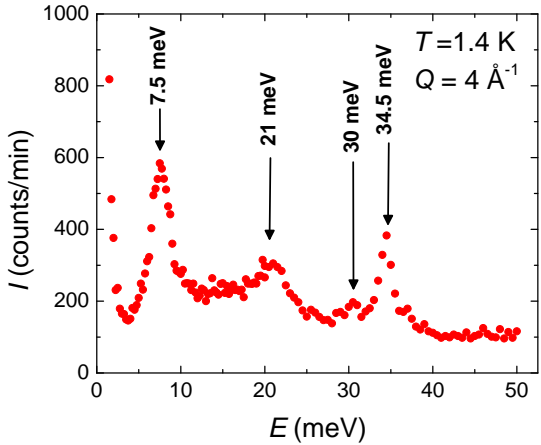}
\begin{center}
\caption{(Color online) Inelastic neutron scattering spectrum 
of 7~g of \cpd\ powder, recorded at 1.4~K
with a wave vector of 4 \AA$^{-1}$. Excited states are found at 7.5, 21
and 30 meV. The sharp peak at 34.5 meV is spurious.}\label{CEF}
\end{center}
\end{figure}

Inelastic neutron scattering measurements provide an independent determination 
of the zero-field CEF splitting, and the results are shown in Fig.~\ref{CEF}. 
We see that there are four sharp peaks, occuring at 7.5 meV, 21 meV, 30 meV, 
and 34 meV.  We believe that the 34 meV peak is a spurious scattering effect 
and that the other three excitations correspond to transitions among the 
CEF-split doublets. While we do not present the data here, the wave vector 
dependence of the intensity of these peaks agrees qualitatively with the 
Yb$^{3+}$ magnetic form
factor, while their excitation energies do not vary. What is more, the values
of the excitation energies found in the inelastic neutron scattering experiment 
are in excellent agreement with those found from the specific heat. We conclude 
that the antiferromagnetic order in \cpd\ involves a doublet state which is 
well separated from the higher lying states. The general success of the 
CEF scenario and the relatively large magnitude of the Yb moments found in the 
neutron diffraction measurements are consistent with a description in which the 
Yb moment is localized, and not significantly itinerant.  No significant difference 
was observed in these excitations as we pass into the antiferromagnetic state.
In principle, we might expect that the exchange interaction would split the 
ground doublet and mix these states with higher lying states. However, our 
energy resolution is not sufficient in this experiment to resolve this effect,
since this splitting should be comparable to \TN=2.4 K, $\sim$ 0.21 meV.


\begin{figure}[!ht]
\includegraphics[width=0.48\textwidth,trim=0in 0in 0in 0in]{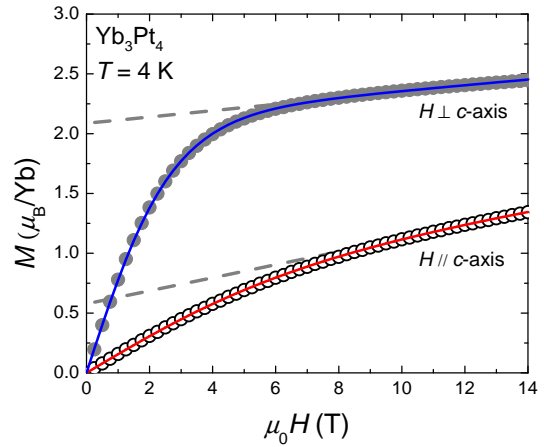}
	\begin{center}
	\caption{(Color online) Magnetization of \cpd\ measured at 4 K for fields $H\bot$ \ca\
		(closed symbols) and for $H //$ \ca\
		(open symbols). The blue line is a
		fit for H$\bot$ \ca, while the red line is the fit for $H //$ \ca, 
		as described in the text. The gray lines are
		extrapolations of the high field magnetization to zero field.\label{MH4K}}
	\end{center}
\end{figure}

A more practical way to extract information about the relative roles of the exchange 
and crystal field anisotropies is from the analysis of the isothermal magnetization. 
$M(H)$ is shown in Fig.~\ref{MH4K} for $T$ = 4 K and for fields as large as 14 T applied 
both parallel and perpendicular to the \ca. 
In both cases, the magnetization is nonlinear, and
resembles a Brillouin function. The magnetization for $H$ $\bot$ \ca\ saturates above $\sim$ 
8 T and the saturation moment is estimated to be 2.12 \mub/Yb, by extrapolating the 
magnetization measured between 8 T and 14 T to zero field, denoted by a dashed gray 
line in Fig.~\ref{MH4K}. The magnetization for $H //$ \ca\ is lower in
all fields, approaching saturation above
$\sim$ 10 T. The saturation moment obtained from extrapolation
is $\sim$ 0.57 \mub/Yb for this field direction.
We use the model that was given by Bonville et al.
~\cite{Bonville92} to describe these magnetization data. This molecular-field model, 
which is solved self-consistently, assumes that the measured magnetization
is generated by the splitting of a well-separated Yb$^{3+}$ doublet ground state 
in an external field, and includes anisotropic $g$ factors, as well as
anisotropic molecular exchange-interaction parameters $\lambda$, and
van Vleck-like terms $\epsilon$ to describe the magnetization which ensues
due to the quantum mechanical mixing of the ground state doublet with the
excited doublets. 
The best fit generated from this model is compared to the $H\bot$ 
\ca\ $M(H)$ curve in Fig.~\ref{MH4K}. We see that the agreement is 
excellent and indicates that $g$=4.23, $\lambda$=0.053~T/\mub,
and $\epsilon$=2.35~$\cdot 10^{-2}$~\mub/T for fields in the \ab.
In the absence of a Van Vleck-like term, this $g$-factor 
would lead to a saturation magnetization of 2.12~\mub/Yb, in
excellent agreement with the zero-field value obtained by extrapolation.
Since the molecular
field parameter $\lambda$ for $H \bot$ \ca\ is weakly positive, we conclude that
exchange interactions in the plane are weak and ferromagnetic.
The analysis is less satisfactory for axial fields $H //$ \ca. A free-parameter fit (red) for the
magnetization with $H$ axial gives $g$=1.44, $\lambda$=1.40~T/\mub,
and $\epsilon$=4.12$\cdot 10 ^{-2}$~\mub/T. 
In the absence of a Van Vleck-like term, this $g$-factor 
would lead to a saturation magnetization of 0.72~\mub/Yb, in
poor agreement with the zero-field value of 0.57~\mub/Yb 
obtained by extrapolation.
In fact, it was not possible to extract a unique set of model parameters from M(H) 
for $H //$\ca, so we cannot comment 
quantitatively on the anisotropy in the 
$g$-factors and the molecular exchange $\lambda$.
Nonetheless, our analysis of the magnetization suggests that the 
CEF-induced magnetic anisotropy strongly favors the crystallographic plane, 
and from the weak molecular exchange found in the \ab, we conclude that the ordering exchange
is anisotropic and favors the \ca.
We note here, that similar, though less conclusive, ideas were presented to describe the magnetic structure 
of isostructural Yb$_{3}$Pd$_{4}$~\cite{Bonville94}.

\section{Conclusions}

We confirm by neutron powder diffraction experiments that the intermetallic compound \cpd\ 
is antiferromagnetic. The zero-field magnetic structure has the same unit cell as the crystal 
structure, so the propagation vector \kvec. The order parameter,  
determined from the temperature dependence of a single-crystal [110] reflection, confirms the 
magnetic ordering temperature at \TN=2.4~K, and indicates that
the magnetic ordering transition is continuous and mean-field like. 
The zero-field magnetic structure was determined using representation analysis.
The fundamental building blocks of the magnetic and crystal structure 
are Yb octahedra, where triads 
of moments are oriented at 120 degrees in the \ab, half with a component 
along the \ca\ and half in the opposite direction.
Because \kvec, there is no net moment within the octahedron.
Individual Yb moments are oriented towards their nearest neighbors,  which lie in
different octahedra. Consequently, the moments have both axial and planar components.  

Like the magnetic-order parameter, the temperature dependence of the specific heat $C(T)$, 
displays a mean-field like transition near \TN=2.4 K. The specific heat step $\Delta C$(\TN) 
amounts to only $\sim$80$\%$ of the value expected for an effective-$S=1/2$ moment. 
It is possible that this reduced anomaly results 
from a substantial degree of hybridization between the Yb moments and the conduction electrons, 
although there are no overt indications of this hybridization, such as an incipient Kondo effect or
any enhancement of the Sommerfeld coefficient $\gamma$, which is found to be less than 
10 mJ/mol K$^{2}$ in the ordered state of \cpd.~\cite{Bennett08}

A primary motivation for our experiments was to determine whether the 
Yb moments can be considered spatially localized or significantly itinerant. 
Neutron diffraction measurements assign an ordered state moment of 
0.81 \mub\ to each Yb ion at 1.4~K, 
which is strongly suggestive that a localized-moment description is more 
suitable. In agreement, specific heat and inelastic neutron scattering 
measurements concur that the moments which order antiferromagnetically 
belong to a well separated doublet ground state, signalling that the 
degeneracy of Yb$^{3+}$ is fully lifted in the low-symmetry crystal 
electric fields experienced by the Yb ions in \cpd. Our analysis of 
the field and temperature dependent magnetization $M(H,T)$ rationalizes 
the magnetic structure, indicating that the dual actions of the crystal
electric field, which seeks to keep the moments in the \ab\ and a 
molecular exchange field which is almost parallel to \ca, but with 
a small transverse component, are together responsible for the canting 
of the Yb moments out of the \ab, while preventing their complete 
alignment along the \ca. The overall success of this crystal field plus exchange 
model is a strong indication that the Yb moments can be considered to be 
spatially localized, and that the f-electrons are excluded from the 
Fermi surface, at least in zero field.

\section{Acknowledgments} \label{ack} We are indebted to A. S. Wills,
A. Kreyssig, P. Stephens, A. Moodenbaugh, W. Ratcliff, and P. Khalifah for valuable discussions
and for help with experiments. Work at the Brookhaven National Laboratory
was supported by the U.S. Department of Energy, Office of Basic Energy Sciences.
Work at Stony Brook University was supported by the National Science
Foundation under grant NSF-DMR-0405961.
We acknowledge the support of the National Institute of Standards and Technology,
U.S. Department of Commerce, in providing the neutron research facilities used in this work.
The identification of any commercial product or trade name does not imply endorsement or recommendation by the National Institute of Standards and Technology.


\begin{thebibliography}{33}
\bibitem[{Lu3()}]{Lu3Pt4}
\bibinfo{note}{Sample was obtained by alloying constituent elements in an
  Ar-atmosphere arc melting furnace.}
\end{thebibliography}
\end{document}